\documentclass[letterpaper]{article} 
\usepackage{aaai2026}  
\usepackage{times}  
\usepackage{helvet}  
\usepackage{courier}  
\usepackage[hyphens]{url}  
\usepackage{graphicx} 
\usepackage[numbers]{natbib}  
\usepackage{caption} 
\usepackage{algorithm}
\usepackage{algorithmic}

\usepackage{newfloat}
\usepackage{listings}
\DeclareCaptionStyle{ruled}{labelfont=normalfont,labelsep=colon,strut=off} 
\floatstyle{ruled}
\newfloat{listing}{tb}{lst}{}
\floatname{listing}{Listing}
\usepackage{svg}
\usepackage{booktabs}
\usepackage{multirow}
\usepackage{amsmath}
\usepackage{amssymb}
\usepackage{dsfont}
\usepackage{subcaption}
\usepackage{hyperref}

\title{Extracting Overlapping Microservices from Monolithic Code via Deep Semantic Embeddings and Graph Neural Network-Based Soft Clustering}
\author {
    Morteza Ziabakhsh\textsuperscript{\rm 1},
    Kiyan Rezaee\textsuperscript{\rm 1},
    Sadegh Eskandari\textsuperscript{\rm 1},
    Seyed Amir Hossein Tabatabaei\textsuperscript{\rm 1},
    Mohammad M. Ghassemi\textsuperscript{\rm 2}
}
\affiliations {
    \textsuperscript{\rm 1}Department of Computer Science, University of Guilan\\
    \textsuperscript{\rm 2}Department of Computer Science and Engineering, Michigan State University\\
    morteza24mail@protonmail.com, kiyanrezaee17@gmail.com, eskandari@guilan.ac.ir, amirhossein.tabatabaei@guilan.ac.ir, ghassem3@msu.edu
}

\begin{document}

\maketitle

\begin{abstract}
Modern software systems are increasingly shifting from monolithic architectures to microservices to enhance scalability, maintainability, and deployment flexibility. Existing microservice extraction methods typically rely on hard clustering, assigning each software component to a single microservice. This approach often increases inter-service coupling and reduces intra-service cohesion. We propose Mo2oM (Monolithic to Overlapping Microservices), a framework that formulates microservice extraction as a soft clustering problem, allowing components to belong probabilistically to multiple microservices. This approach is inspired by expert-driven decompositions, where practitioners intentionally replicate certain software components across services to reduce communication overhead. Mo2oM combines deep semantic embeddings with structural dependencies extracted from method-call graphs to capture both functional and architectural relationships. A graph neural network-based soft clustering algorithm then generates the final set of microservices. We evaluate Mo2oM on four open-source monolithic benchmarks and compare it against eight state-of-the-art baselines. Our results demonstrate that Mo2oM achieves improvements of up to 40.97\% in structural modularity (balancing cohesion and coupling), 58\% in inter-service call percentage (communication overhead), 26.16\% in interface number (modularity and decoupling), and 38.96\% in non-extreme distribution (service size balance) across all benchmarks.
\end{abstract}

\begin{links}
    \link{Code}{https://github.com/Morteza-24/Monolith-to-Microservice}
    \link{Datasets}{https://github.com/Morteza-24/Monolith-to-Microservice/tree/main/test_projects}
\end{links}

\section{Introduction}

\begin{figure}[t]
    \centering
    \includegraphics[width=\linewidth]{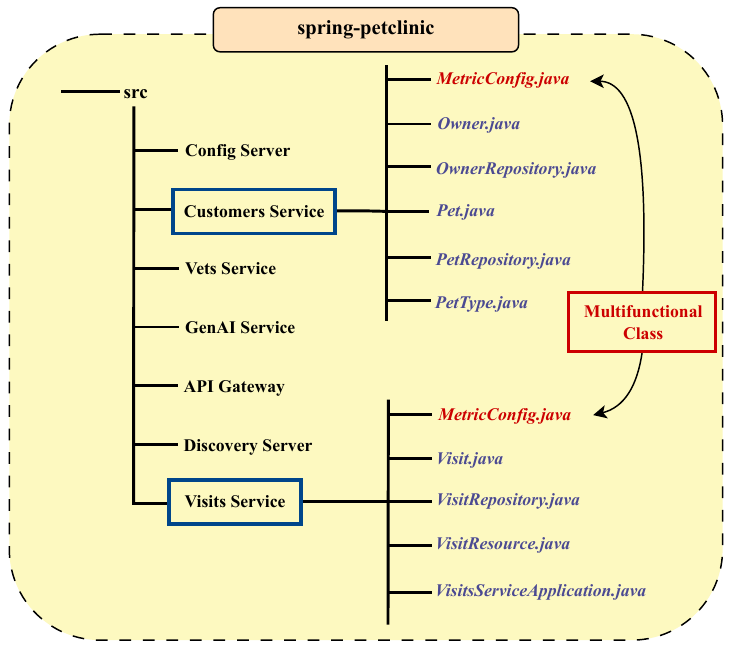}
    \caption{\textbf{Our Motivation.} Example of overlapping class memberships in the Spring Petclinic\protect\footnotemark microservices architecture. The \texttt{MetricConfig} class is shared across the Customers and Visits microservices to support cross‐cutting metrics functionality, illustrating the need for soft clustering rather than strict partitioning.}
    \label{fig:metricconfig}
\end{figure}

Microservices architecture has emerged as the de facto standard for building modern software systems. This approach decomposes a single application into a suite of small, independently deployable services, each encapsulating a specific business capability and communicating via lightweight protocols.  microservices architecture enables fine-grained scaling, cloud-native integration, and strong fault isolation, qualities that are essential for large-scale systems in real-world applications like e-commerce, real-time analytics, and finance ~\cite{desina2023evaluating, abgaz2023decomposition, barzotto2022evaluation, oyeniran2024microservices, seedat2023systematic, henriquez2025architectural, katal2025evolution}. However, many legacy systems remain monolithic – characterized by a single, tightly-coupled codebase where all components are developed, deployed, and scaled as one unit. Within such monoliths, scaling and maintenance become increasingly difficult, and a single fault can jeopardize the entire application~\cite{10763474, 10475769, su2024microservice, 10431792, soylemez2022challenges}.\footnotetext{\url{https://github.com/spring-petclinic/spring-petclinic-microservices}} Organizations often opt to restructure monoliths into microservices by identifying service candidates within existing codebases to mitigate these issues. This strategy is preferred over full rewrites due to the substantial risk, time, and resource investment required to rebuild complex business logic from scratch, which can disrupt operations and delay value realization. The effective migration from monolithic to microservices architecture is therefore a critical challenge in modern software engineering, enabling organizations to unlock agility, resilience, and scalability while preserving existing investments and business knowledge embedded in legacy systems.

The migration from monolithic architectures to microservices has generated extensive research, resulting in numerous approaches proposed for identifying service boundaries within legacy systems. State-of-the-art microservice extraction methods typically follow a two-step workflow. First, they perform \textit{dependency analysis}, mining semantic and structural relationships between code elements using techniques such as TF-IDF, cosine similarity, or call-graph analysis. Next, they apply \textit{clustering} algorithms, such as DBSCAN, to group components into services~\cite{sellami2022hierarchical, mazlami2017extraction, scanniello2010approach}. However, a key limitation of these approaches is that they enforce hard clustering, assigning each class to exactly one microservice. This rigid partitioning forces multifunctional software components into a single microservice. Consequently, it generates excessive inter-service calls, reduces cohesion (the internal consistency within a microservice), and increases coupling (the interdependencies between microservices)~\cite{kalske2017transforming, desai2021graph, mazlami2017extraction}.

Interestingly, our analysis of expert-driven decompositions in real‑world systems reveals that microservice boundaries often overlap. For instance, in the manually refactored Spring Petclinic microservices application, the \texttt{MetricConfig} class is shared between two microservices to support cross‑cutting functionality (see Fig.~\ref{fig:metricconfig}). This finding leads us to our central question: \textit{Could permitting multifunctional classes to span multiple microservices actually improve the quality of microservices-based systems?}

Meanwhile, the rapid advancement of large language models (LLMs) for code understanding \cite{zhou2025large,hou2024large,xu2022systematic} presents new opportunities to revolutionize microservice extraction. However, current microservice extraction techniques still rely on traditional feature extraction methods. As a result, they fundamentally lack the ability to capture the rich semantic relationships and functional intent embedded in source code. For example, two classes implementing different aspects of authentication, such as token validation and password encryption, may exhibit minimal structural dependencies, but share deep semantic connections that would naturally group them within the same microservice. This semantic gap in existing approaches often leads to suboptimal decompositions, where functionally related classes are separated across services. This observation leads us to our second key research question: \textit{Could leveraging code-LLM generated semantic embeddings provide more functionally coherent microservice decompositions compared to traditional feature extraction methods?}

To address these, we propose \textbf{Mo2oM} (\underline{Mo}nolithic \underline{to} \underline{o}verlapping \underline{M}icroservices), the first framework to frame microservice extraction as a \textit{soft clustering} problem. Rather than enforcing binary assignments, Mo2oM assigns each class a membership score across all microservices. Our approach comprises three interrelated components. First, we utilize semantic-aware code embeddings generated by a large language model fine-tuned on source code, ensuring that the extracted vectors accurately capture functional and semantic similarities among code elements. Second, we combine these embeddings with structural information derived from method-call graphs to represent logical interactions and architectural dependencies within the codebase. Finally, we implement a graph neural network (GNN)-based soft clustering scheme in which each class is probabilistically assigned to microservices, forming a membership matrix. The final set of microservices is obtained by applying a threshold to this matrix, assigning each class to all services where its membership value exceeds the threshold.

We evaluate Mo2oM on several large monolithic systems and benchmark its performance against strong baselines using standard metrics. Our results demonstrate that across all benchmarks, Mo2oM consistently boosts intra-service cohesion and reduces inter-service coupling, delivering up to a 48.4\% overall improvement over the best hard-clustering competitor.

In summary, our main contributions are as follows:

\begin{itemize}
\item To the best of our knowledge, we propose the first soft clustering approach for microservice extraction, allowing classes to have overlapping microservice memberships.
\item We combine LLM-driven semantic embeddings with structural dependency analysis to achieve a more comprehensive modeling of class relationships.
\item We perform thorough ablation studies and hyperparameter analysis to assess the contribution of each component in our framework.
\item We conduct extensive empirical evaluations of Mo2oM, showing consistent improvements over leading state-of-the-art methods.
\end{itemize}


The remainder of this paper is organized as follows. Section~2 reviews related work, categorizing existing decomposition techniques into dynamic and static approaches, and identifies key limitations. Section~3 presents our proposed approach. Section~4 provides a comprehensive evaluation of Mo2oM against state-of-the-art baselines, along with an ablation study quantifying the impact of each component. Finally, Section~5 concludes the paper.

\section{Related Work}\label{related}

Software decomposition techniques for migrating monolithic systems to microservices are broadly categorized into dynamic and static analysis approaches. Dynamic approaches use runtime information, whereas static methods rely on source code-level information. We review representative methods in each category, highlight open challenges, and position our Mo2oM method within the research landscape.

\subsection{Dynamic Analysis Approaches}

Dynamic methods collect execution traces to capture invocation patterns that static analysis may miss, such as reflective calls, dynamic class loading, or dependency injection. Mono2Micro~\cite{kalia2021mono2micro} records call sequences of a Java application under varied workloads, builds spatio-temporal graphs of class interactions, and applies clustering to derive candidate microservices. FoSCI~\cite{jin2019service} similarly uses the Kieker~\cite{van2012kieker} tracing framework to log method executions across functional test suites and groups classes by their co-occurrence in traces. While runtime-based techniques are effective at capturing actual dependency usage, they incur high overhead due to trace collection and analysis. Moreover, their accuracy depends heavily on comprehensive test coverage—unexercised execution paths can result in missed dependencies.

\subsection{Static Analysis Approaches}

Static analysis methods infer relationships from source code structure, comments, and version history, offering scalability and applicability without runtime data. Early work, inspired by Parnas's information-hiding principles~\cite{parnas1972criteria}, used coupling and cohesion metrics to decompose systems, as seen in tools like ARCH and ACDC~\cite{schwanke1991intelligent,tzerpos2000accd}.

Graph-based techniques model classes as nodes and relationships (e.g., method calls, inheritance) as edges. ServiceCutter employs graph-cutting algorithms, such as Girvan-Newman and epidemic label propagation, using sixteen coupling criteria to identify microservice boundaries~\cite{gysel2016service}. MEM enhances this by incorporating semantic similarity from identifier names and comments, as well as contributor-based coupling from commit history~\cite{mazlami2017extraction}. Recent advancements leverage Graph Neural Networks (GNNs). For instance, CoGCN constructs heterogeneous graphs with structural, semantic, and historical attributes, applies a two-layer GCN to generate class embeddings, and clusters them into microservices~\cite{desai2021graph}.

Search- and density-based clustering methods treat service extraction as an optimization problem. The Bunch tool~\cite{mancoridis1999bunch} and simulated annealing approaches~\cite{koschke2000evaluation} use genetic algorithms and hill climbing to navigate the decomposition space. HyDec~\cite{sellami2022hierarchical} integrates structural similarity from call-graph analysis and semantic similarity via TF-IDF and cosine distance, applying hierarchical density-based clustering to propose microservice candidates.

\subsection{Advances in Clustering Schemes}

Recent methods explore more flexible clustering algorithms to improve decomposition. GCD-DVF~\cite{qian2023microservice}, a dynamic approach, combines structural dependencies and runtime traces using a graph attention adaptive residual network for embedding and utilizes a Fuzzy C-Means (FCM) network with an argmax function for clustering, resulting in hard partitioning. DEEPLY~\cite{yedida2023expert}, a static approach, builds upon the CoGCN algorithm and enhances it by incorporating a weighted loss function and hyper-parameter tuning via the Tree of Parzen Estimator (TPE), using spectral clustering for flexible cluster shapes. These advances highlight a research trend toward more flexible and fuzzy clustering algorithms for microservice extraction.

\subsection{Limitations and Our Contribution}

Most existing approaches rely on traditional methods like TF-IDF to represent class semantics, which often fail to capture deeper contextual relationships. Pre-trained transformers have recently shown superior semantic embeddings for source code~\cite{petridis2024text,guo2022unixcoder}. Furthermore, all existing clustering schemes enforce hard partitioning, where each class belongs to exactly one microservice, leading to increased inter-service communication and coupling~\cite{kalske2017transforming, desai2021graph, mazlami2017extraction}. Our work addresses these limitations by integrating transformer-based embeddings with a flexible, overlapping clustering framework that allows classes to participate in multiple microservices where appropriate.

\section{Monolithic to Overlapping Microservices (Mo2oM)}\label{proposed}

\begin{figure*}[t]
    \centering
    \includegraphics[width=0.9\linewidth]{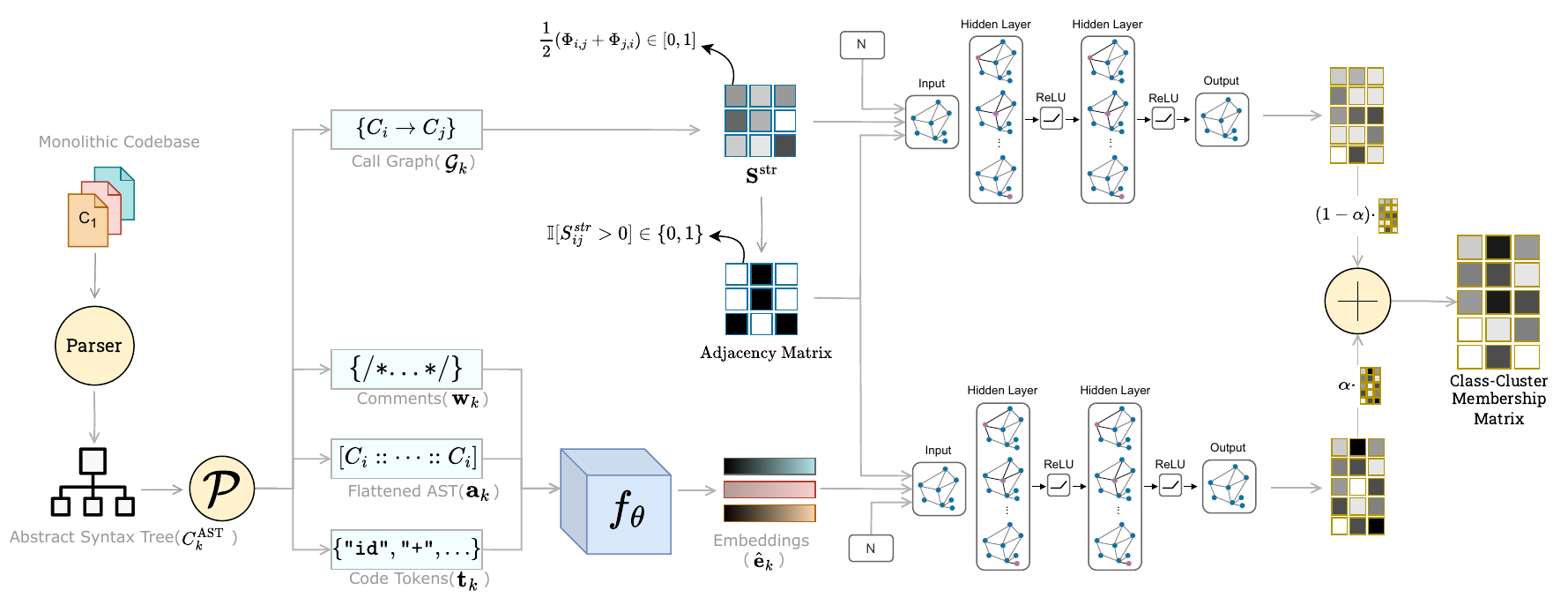}
    \caption{\textbf{Overview of the Mo2oM Decomposition Framework.} The pipeline involves parsing the monolithic application, extracting structural and semantic features from classes, and applying soft clustering to generate candidate microservices.}
    \label{fig:mainfig_abstract}
\end{figure*}

The migration from a monolithic architecture to a microservices-based system involves decomposing a complex, unified codebase into smaller, independent services. Formally, given a monolithic codebase comprising a set of $Y$ classes $\{C_i\}_{i=1}^Y$ and a target set of $N$ microservices $\{M_j\}_{j=1}^N$, the objective is to optimally assign classes to microservices such that inter-service dependencies are minimized while intra-service cohesion is maximized. Our proposed method, Mo2oM, addresses this challenge by first extracting both structural and semantic features from each class, the smallest unit of decomposition in object-oriented systems. These features then serve as inputs to a soft-clustering algorithm that generates overlapping candidate microservices, allowing classes to be included in multiple services where appropriate. An overview of the Mo2oM framework is illustrated in Figure~\ref{fig:mainfig_abstract}.

\subsection{Parsing and Structural Artifact Extraction}\label{subsec:parsing}

We begin the decomposition of a monolithic codebase by parsing each class $C_k$ into its abstract syntax tree (AST) representation $C^{\text{AST}}_k$. We define a structural artifact extraction function:
\begin{equation}
    \mathcal{P}: C^{\text{AST}}_k \rightarrow \langle \mathcal{G}_k, \mathbf{t}_k, \mathbf{w}_k, \mathbf{a}_k \rangle
\end{equation}
which maps the AST to a tuple of four structural components. Here, $\mathcal{G}_k = (V_k, E_k, I_k)$ represents the intra-class call graph, where $V_k$ is the set of method nodes, $E_k$ denotes intra-class method invocations, and $I_k$ captures inter-class dependencies through outgoing calls. The vector $\mathbf{t}_k$ represents the sequence of source code tokens extracted from $C_k$, $\mathbf{w}_k$ contains code comments associated with $C_k$, and $\mathbf{a}_k = \mathcal{F}(C^{\text{AST}}_k)$ is the flattened AST obtained via a bijective transformation $\mathcal{F}$, which linearizes the hierarchical AST into a sequential representation of syntactic subgraphs (Algorithm~\ref{alg:ast_flattening})~\cite{guo2022unixcoder}. These components serve as the foundation for subsequent structural and semantic feature extraction in the Mo2oM framework.

\begin{algorithm}[t]
\caption{AST Flattening ($\mathcal{F}$)}
\label{alg:ast_flattening}
\begin{algorithmic}[1]
\REQUIRE AST node $n$ \COMMENT{Current subtree root}
\STATE Initialize $\mathbf{a}_k \gets []$
\IF{$n$ is a leaf node}
    \STATE $\mathbf{a}_k \gets \mathbf{a}_k \oplus n.\text{name}$
\ELSE
    \STATE $\mathbf{a}_k \gets \mathbf{a}_k \oplus n.\text{name}::\text{left}$
    \FOR{each child $c$ of $n$}
        \STATE $\mathbf{a}_k \gets \mathbf{a}_k \oplus \mathcal{F}(c)$
    \ENDFOR
    \STATE $\mathbf{a}_k \gets \mathbf{a}_k \oplus n.\text{name}::\text{right}$
\ENDIF
\RETURN $\mathbf{a}_k$
\end{algorithmic}
\end{algorithm}

\subsection{Structural Dependency Features}\label{subsec:structural_features}

To uncover latent coupling and communication patterns within a monolithic codebase, we analyze structural dependencies based on method invocation patterns between classes. This analysis identifies communication hotspots, where frequent method calls suggest classes should be co-located within the same microservice to minimize latency. Additionally, it highlights potential performance bottlenecks arising from excessive cross-service interactions, which could degrade system efficiency. These structural features capture dependencies and collaboration among classes.

For structural analysis, two classes \(C_i\) and \(C_j\) are considered dependent if their methods frequently invoke one another. Let \(\phi_{i \rightarrow j} = \left|\{ (v_i, v_j) \in I_i \mid v_i \in V_i, v_j \in V_j \}\right|\) represent the number of method calls from \(C_i\) to \(C_j\), recorded in \(\mathcal{G}_i\). To account for varying call frequencies, we normalize these counts by the total incoming calls to \(C_j\) (in-degree), defined as \(\psi_j = \sum_x \phi_{x \rightarrow j}\). The normalized influence of \(C_i\) on \(C_j\) is then computed as \(\Phi_{i,j} = \phi_{i \rightarrow j} / \psi_j\). Finally, we construct a symmetric structural dependency matrix \(\mathbf{S}^{\mathrm{str}}\) by averaging the normalized influences in both directions:
\begin{equation}
\mathbf{S}^{\mathrm{str}}(i,j) = \frac{1}{2} \left( \Phi_{i,j} + \Phi_{j,i} \right) \in [0,1]
\end{equation}
where values closer to 1 indicate stronger mutual dependencies between \(C_i\) and \(C_j\).

\subsection{Semantic Features}\label{subsec:semantic}

While structural dependencies capture the interactions between classes, semantic features encode the intrinsic functionality and intent of each class. To capture functional semantics aligned with business capabilities, we employ multimodal representation learning to embed class-level information into a latent space. Each class \( C_k \) is encoded through two complementary modalities: lexical features \( \mathbf{x}_k^{\text{lex}} = (\mathbf{t}_k, \mathbf{w}_k) \) comprising tokenized source code and natural language documentation, and syntactic features \( \mathbf{x}_k^{\text{AST}} = \mathbf{a}_k \) derived from the linearized structural representation. These modalities are fused through a transformer encoder \( f_\theta \) to generate normalized latent representations according to:
\begin{equation}
    \hat{\mathbf{e}}_k = \frac{f_\theta(\mathbf{x}_k^{\text{lex}}, \mathbf{x}_k^{\text{AST}})}{\|f_\theta(\mathbf{x}_k^{\text{lex}}, \mathbf{x}_k^{\text{AST}})\|_2}
\end{equation}
where \( \hat{\mathbf{e}}_k \in \mathbb{R}^d \) denotes the unit-normalized embedding. We implement \( f_\theta \) using UniXcoder~\cite{guo2022unixcoder} for its demonstrated efficacy in cross-modal fusion for software understanding. The resulting embeddings distill syntactic patterns and semantic intent into compact geometric representations suitable for the downstream clustering task.

\subsection{Soft Clustering}\label{subsec:nocd}

To generate soft cluster assignments for microservices, we apply two instances of the Neural Overlapping Community Detection (NOCD) algorithm~\cite{shchur2019overlapping}: one driven by structural dependencies and the other by semantic embeddings. Each NOCD variant takes as input a corresponding feature matrix, and a shared adjacency matrix and produces a class–cluster membership matrix. The shared adjacency matrix is defined as: 
\begin{equation}
A_{ij} = \mathds{1}[\mathbf{S}^{\mathrm{str}}_{ij} > 0]
\end{equation}
where $\mathds{1}[\cdot]$ is the indicator function.

\textbf{Structural NOCD} optimizes a two-layer graph neural network over $A$ and $\mathbf{S}^{str}$ as node features, yielding a membership matrix:
\begin{equation}
\begin{split}
M^{\mathrm{str}} &= \mathrm{NOCD}(S^{\mathrm{str}}, A, N) \\
&= \text{ReLU}\Big( \widetilde{A} \cdot \text{ReLU}\big(
\widetilde{A} \, \mathbf{S}^{\mathrm{str}}\,W_1
\big) W_2 \Big) \in \mathbf{R}^{Y \times N}
\end{split}
\end{equation}
where $\widetilde{A} = D^{-1/2}AD^{-1/2}$ is the normalized adjacency matrix, and $D_{ii}=\sum_{j}A_{ij}$ is the diagonal degree matrix of $A$. Moreover, $N$ is the number of target microservices.

\textbf{Semantic NOCD}  applies NOCD on features $E = [\hat{\mathbf e}_1, \dots, \hat{\mathbf e}_Y]^\top \in \mathbb{R}^{Y\times d}$:
\begin{equation}
M^{\mathrm{sem}} = \mathrm{NOCD}(E, A, N) \in \mathbf{R}^{Y \times N}.
\end{equation}

Both models are trained by minimizing the negative log-likelihood of a Bernoulli-Poisson edge model:
\begin{equation}
\begin{split}
\mathcal{L}^u &= - \sum_{(i,j) \in \mathcal{E}} \log \left( 1 - \exp\left( - \langle M^{\mathrm{u}}_i, M^{\mathrm{u}}_j \rangle \right) \right)\\
&+ \sum_{(i,j) \in \mathcal{N}} \langle M^{\mathrm{u}}_i, M^{\mathrm{u}}_j \rangle
\end{split}
\end{equation}
where $u \in \{\mathrm{str},\mathrm{sem}\}$, $\mathcal{E} \subseteq \{(i,j) \mid A_{ij} = 1\}$ denotes the set of observed edges in the graph, and $\mathcal{N} \subseteq \{(i,j) \mid A_{ij} = 0\}$ is a set of randomly sampled non-edge pairs used for negative sampling. The dot product $\langle M^{\mathrm{u}}_i, M^{\mathrm{u}}_j \rangle$ captures the similarity between nodes $i$ and $j$ in the embedding space, promoting high affinity for connected nodes and penalizing spurious similarity between unconnected ones.

We compute the final microservice affiliation matrix by fusing the structural and semantic membership matrices:
\begin{equation}
\label{eq:f-combined}
M = \alpha M^{sem} + (1 - \alpha) M^{str}, \quad \alpha \in [0,1]
\end{equation}
where $\alpha$ is a hyperparameter to control the tradeoff between semantic and structural features. Class $C_i$ is assigned to $j$th microservice when $M_{ij} \geq \tau$, with threshold $\tau \in (0,1)$ controlling assignment strictness. Classes with all $M_{ij} < \tau$ are flagged as outliers.

\section{Experiments}\label{experiments}

In this section, we evaluate the effectiveness of our proposed Mo2oM approach. We begin by detailing the benchmarks, baselines and evaluation metrics used, followed by implementation specifics. Next, we present a comparative analysis against state-of-the-art methods to demonstrate performance improvements. An ablation study then examines the contributions of individual components and analyzes the sensitivity of the parameters. Finally, we conclude with a discussion on the practical applicability of our method in refactoring existing microservices.

\begin{table*}[t]
\renewcommand{\arraystretch}{1.5}
\resizebox{\textwidth}{!}{\begin{tabular}{lcccc|cccc|cccc|cccc}
\toprule
\multirow{2}{*}[-2.5pt]{\textbf{Method}} & \multicolumn{4}{c}{\textbf{JPetStore}} & \multicolumn{4}{c}{\textbf{DayTrader}} & \multicolumn{4}{c}{\textbf{AcmeAir}} & \multicolumn{4}{c}{\textbf{Plants}} \\
\cmidrule[0.5pt]{2-17}
 & \textbf{SM $\uparrow$} & \textbf{ICP $\downarrow$} & \textbf{IFN $\downarrow$} & \textbf{NED $\downarrow$} & \textbf{SM  $\uparrow$} & \textbf{ICP $\downarrow$} & \textbf{IFN $\downarrow$} & \textbf{NED $\downarrow$} & \textbf{SM $\uparrow$} & \textbf{ICP $\downarrow$} & \textbf{IFN $\downarrow$} & \textbf{NED $\downarrow$} & \textbf{SM $\uparrow$} & \textbf{ICP $\downarrow$} & \textbf{IFN $\downarrow$} & \textbf{NED $\downarrow$} \\
\midrule
Bunch & 0.111 & 0.264 & 4.250 & \textbf{0.000} & 0.121 & 0.425 & 10.400 & 0.639 & 0.044 & 0.330 & 7.333 & \underline{0.154} & 0.096 & 0.443 & 7.667 & \textbf{0.000} \\
FoSCI & 0.050 & 0.398 & 3.792 & 0.515 & 0.094 & 0.725 & 5.053 & 0.705 & 0.095 & 0.706 & 4.375 & 0.407 & 0.115 & 0.699 & 4.938 & 0.653 \\
\midrule
MEM & 0.121 & 0.434 & 3.429 & 1.000 & 0.089 & 0.355 & 3.357 & 1.000 & 0.097 & 0.589 & 4.333 & 0.464 & 0.210 & 0.320 & 4.750 & 0.288 \\
\midrule
CoGCN & 0.079 & 0.510 & 2.531 & 0.609 & 0.086 & \underline{0.300} & 2.600 & 0.676 & 0.038 & 0.444 & 2.800 & 0.250 & 0.083 & 0.636 & 4.938 & 0.443 \\
DEEPLY & \underline{0.170} & 0.347 & 1.960 & 0.928 & 0.115 & 0.368 & 2.748 & 0.857 & 0.089 & 0.380 & 2.237 & 0.778 & 0.215 & 0.467 & 4.500 & 0.938 \\
GDC-DVF & 0.108 & 0.274 & \textbf{1.375} & 0.351 & 0.117 & 0.329 & 2.625 & 0.419 & 0.094 & 0.290 & 1.667 & 0.321 & 0.152 & 0.370 & 4.125 & 0.231 \\
\midrule
Mono2Micro & 0.052 & 0.318 & 2.322 & 0.216 & 0.084 & 0.346 & 2.421 & \underline{0.338} & 0.072 & 0.527 & 3.625 & 0.429 & 0.078 & 0.381 & 6.000 & \underline{0.038} \\
HyDec & 0.035 & \textbf{0.037} & 3.000 & 1.000 & \underline{0.498} & 0.424 & \underline{1.375} & 0.922 & \underline{0.227} & \underline{0.198} & \underline{1.600} & 0.863 & \underline{0.509} & \underline{0.300} & \underline{2.286} & 1.000 \\
\midrule
Mo2oM (ours) & \textbf{0.288} & \underline{0.153} & \underline{1.414} & \underline{0.118} & \textbf{0.621} & \textbf{0.153} & \textbf{1.118} & \textbf{0.225} & \textbf{0.337} & \textbf{0.100} & \textbf{1.286} & \textbf{0.094} & \textbf{0.692} & \textbf{0.126} & \textbf{1.688} & \textbf{0.000} \\
\bottomrule
\end{tabular}}
\caption{Comparative Performance of Mo2oM and Baseline Methods Across Multiple Benchmarks. \textbf{Bold}: best. \underline{Underline}: runner-up. Arrows indicate whether lower (↓) or higher (↑) values are better.}
\label{tab:result}
\end{table*}

\subsection{Setup}\label{ex:setup}

\paragraph{Benchmarks}
We evaluate our approach on four widely used open-source monolithic Java applications: \textit{JPetStore}\footnote{\url{https://github.com/KimJongSung/jPetStore}}, \textit{DayTrader}\footnote{\url{https://github.com/WASdev/sample.daytrader7}}, \textit{AcmeAir}\footnote{\url{https://github.com/acmeair/acmeair}}, and \textit{Plants}\footnote{\url{https://github.com/WASdev/sample.plantsbywebsphere}}. These applications cover diverse domains and vary significantly in complexity, codebase size, and functional scope, aligning with prior work on monolithic-to-microservices decomposition~\cite{kalia2021mono2micro,sellami2022hierarchical,desai2021graph}. \textit{JPetStore} is an e-commerce application consisting of 73 classes and 372 methods. \textit{DayTrader} is a financial trading platform implemented with 108 classes and 958 methods. \textit{AcmeAir} is a cloud-based airline reservation system comprising 68 classes and 525 methods. Finally, \textit{Plants} is a resource management system with 37 classes and 426 methods.

\paragraph{Baselines}
We compare our method against eight baselines representing diverse approaches. These include methods based on genetic and evolutionary algorithms, Bunch~\cite{mitchell2006automatic} and FoSCI~\cite{jin2019service}); a formal modeling approach, MEM~\cite{mazlami2017extraction}; deep learning methods, including CoGCN~\cite{desai2021graph}, DEEPLY~\cite{yedida2021partitioning}, and GDC-DVF~\cite{qian2023microservice}; and clustering-based approaches, Mono2Micro~\cite{kalia2021mono2micro} and HyDec~\cite{sellami2022hierarchical}. A common constraint across all baselines is the assignment of each class to at most one microservice.

\paragraph{Evaluation Metrics}
We evaluate the quality of generated microservices using four metrics: Structural Modularity (SM) \cite{jin2019service}, Inter-Call Percentage (ICP) \cite{kalia2020mono2micro}, Interface Number (IFN) \cite{jin2019service}, and Non-Extreme Distribution (NED) \cite{desai2021graph,scanniello2010approach}. SM captures how well the system is decomposed by balancing intra-service cohesion and inter-service coupling, with higher values indicating cleaner modular boundaries. IFN measures the complexity of microservices by counting the number of interfaces each service exposes. A lower IFN value indicates a cleaner and more manageable system with fewer dependencies between services.  ICP quantifies the proportion of method calls occurring across services, with lower values reflecting better cohesion and less communication overhead between microservices. NED evaluates how evenly classes are distributed among microservices. A well-balanced system avoids extreme variations in microservice sizes, ensuring that no service is disproportionately large or small. In this study, we use a slightly modified version of NED so that it works in soft clustering scenarios as well. It is computed as follows:
\begin{equation}
\text{NED} = 1 - \frac{\sum_{i=1, \, m_i \in \mathcal{M}_{\text{non-extreme}}}^{N} |m_i|}{\sum_{i=1, \, m_i \in \mathcal{M}}^{N} |m_i|}
\end{equation}
where $\mathcal{M}$ is the set of all microservices, and $\mathcal{M}_{\text{non-extreme}}$ is the set of microservices classified as non-extreme. A microservice $m_i$ is considered non-extreme if $5 \leq |m_i| \leq 20$, as suggested in~\cite{scanniello2010approach}. A lower
NED value signifies a more evenly distributed system.

\paragraph{Implementation Details}
We employ the \textit{JavaParser} library~\cite{vanBruggen2023JavaParser} to construct ASTs. The NOCD algorithm is implemented using a two-layer Graph Convolutional Network (GCN), where each layer consists of 128 hidden units and a dropout rate of 0.5. Training is performed with the Adam optimizer using a learning rate of \(10^{-3}\) and $L2$ regularization. We adopt a batch-wise stochastic loss computation and apply early stopping, terminating training if the validation loss does not improve for 10 consecutive epochs. For the number of clusters, a shared hyperparameter across Mo2oM, MEM, FoSCI, Mono2Micro, CoGCN, and GDC-DVF, we adopt the approach from ~\cite{scanniello2010using} and \cite{kalia2020mono2micro}. Specifically, we evaluate cluster counts ranging from \(\lceil Y/2 \rceil\) down to a minimum of 2 or 3, where \(Y\) denotes the number of classes in the monolithic system, decreasing in steps of 2. For each baseline model, we adopt the hyperparameter configuration recommended by its respective authors.

\subsection{Experimental Results}\label{ex}

We present the comparison between our approach and the baseline methods in Table~\ref{tab:result}.
Regarding SM, our Mo2oM significantly outperforms all other baseline methods by a substantial margin across all four benchmarks, clearly demonstrating that our approach is an effective way for extracting highly cohesive microservices. It obtains 69.4\% improvement on JPetStore compared to the previous SOTA DEEPLY~\cite{yedida2021partitioning}. Additionally, it shows improvements of 24.7\%, 48.5\%, and 35.9\% over DayTrader, AcmeAir, and Plants, respectively, when compared to the previous SOTA HyDec~\cite{sellami2022hierarchical}.

In terms of ICP, Mo2oM demonstrates superior performance on three out of the four benchmarks. It achieves a 49.0\% reduction on DayTrader compared to the strongest baseline, CoGCN~\cite{desai2021graph}. The improvements on AcmeAir and Plants are 49.5\% and 58.0\%, respectively. Although HyDec achieves a lower ICP on JPetStore, this comes at the cost of extremely poor service size balance, as indicated by its NED score of 1.000.

For IFN, which measures the average number of interface classes, Mo2oM delivers strong performance. It reduces IFN by 18.7\% on DayTrader 19.6\% on AcmeAir, and 26.2\% on Plants compared to the SOTA HyDec. On JPetStore, GDC-DVF~\cite{qian2023microservice} achieves a marginally 2.8\% lower IFN than Mo2oM. However, GDC-DVF performs significantly worse across the other three metrics.

Finally, considering NED, Mo2oM consistently extracts the most balanced microservices. It improves this metric by 33.4\% on DayTrader compared to the SOTA Mono2Micro~\cite{kalia2021mono2micro} and 38.9\% on AcmeAir compared to Bunch~\cite{mitchell2006automatic}. Moreover, our algorithm achieves perfect balance on Plants.

\subsection{Ablation Study}\label{sec:ablation}

We conduct an ablation study to isolate the contributions of our LLM-based semantic encoder and soft-clustering mechanism in the Mo2oM method, as well as to assess the sensitivity of key hyperparameters.


\begin{figure*}[t]
  \centering
  \includegraphics[width=\linewidth]{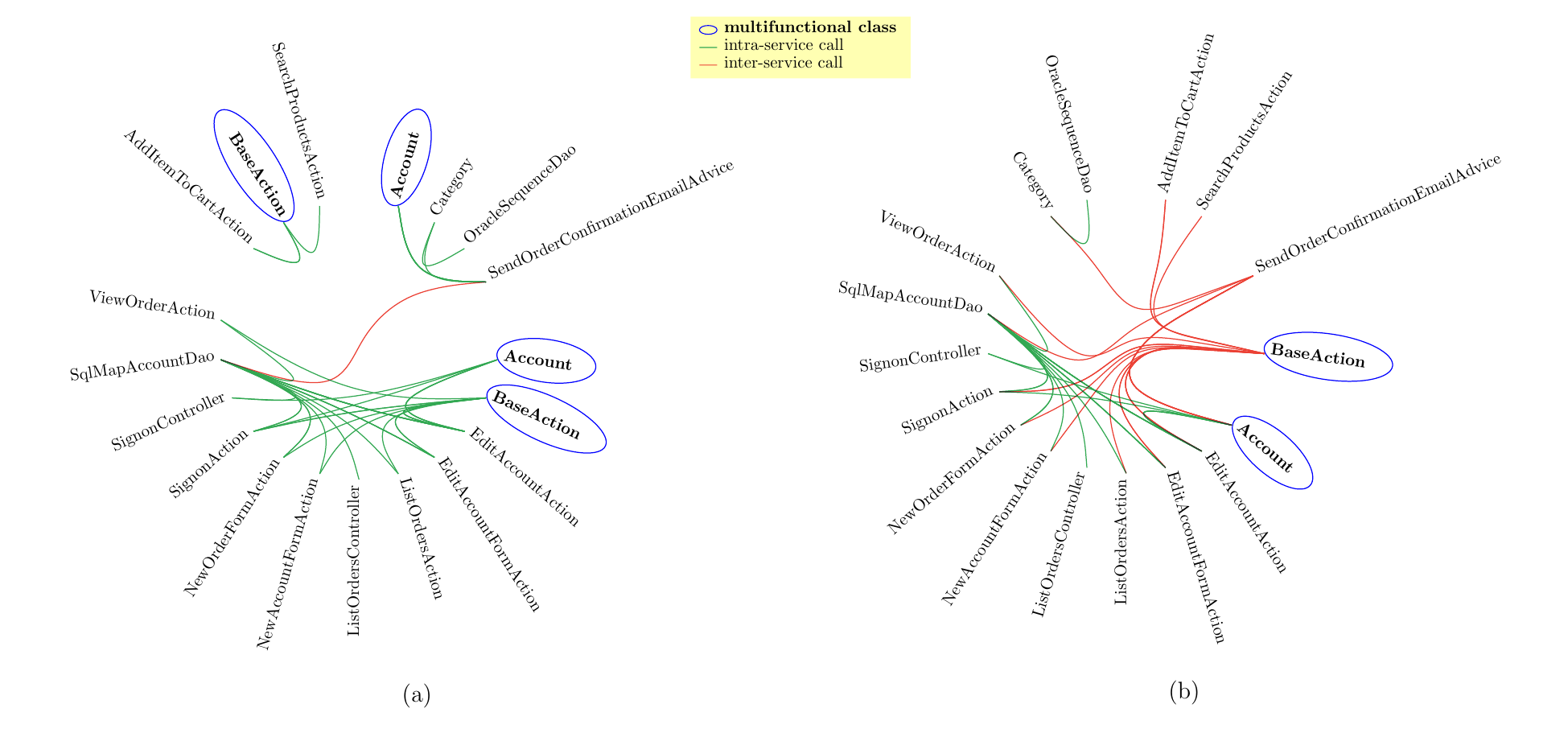}
  \caption{\textbf{Microservice Decomposition with Soft vs. Hard Clustering (JPetStore)}. Comparison of (a) soft clustering (UniXcoder + Soft) and (b) hard clustering (UniXcoder + Hard). Green lines: intra-service calls; red lines: inter-service calls; blue ellipses: multifunctional classes. Soft clustering reduces inter-service dependencies by allowing shared class ownership.}
  \label{fig:chord}
\end{figure*}

\paragraph{Effects of Soft Clustering}
To quantify the benefit of overlapping memberships, we compare our approach (UniXcoder + Soft) against a variant (UniXcoder + Hard) that uses UniXcoder-based semantic and structural similarities but enforces hard clustering by assigning each class to a single microservice based on its highest membership value. Table~\ref{tab:soft-hard} shows that hard clustering significantly degrades performance. For instance, SM decreases by 0.34 on JPetStore, and ICP increases from 0.082 to 0.442 on DayTrader. On average, soft clustering improves SM by 175\%, ICP by 77.1\%, IFN by 40.8\%, and NED by 56.3\%. These results underscore the importance of overlapping assignments in capturing cross-cutting dependencies and ensuring cohesive service boundaries. Figure~\ref{fig:chord} depicts the impact of soft clustering visually on JPetStore: hard clustering forces a strict one-to-one assignment of classes—particularly multifunctional ones like \texttt{Account} and \texttt{BaseAction}—to a single microservice, fracturing natural cohesion and necessitating costly inter-service calls (red lines). In contrast, soft clustering allows these classes to participate in multiple microservices (blue ellipses), preserving strong intra-service cohesion (green lines) while minimizing inter-service dependencies.

\paragraph{Effects of LLM-based Semantic Encoder}
To assess the impact of our deep semantic embeddings, we compare our approach with a variant (TF-IDF + Soft) that replaces UniXcoder embeddings with TF-IDF vectors for semantic similarity, while retaining NOCD soft-clustering. Table~\ref{tab:unixcoder-tfidf} shows that UniXcoder embeddings yield consistent improvements across all benchmarks. While TF-IDF + Soft occasionally achieves higher SM, it increases IFN or NED, indicating less modular and less balanced microservices. On average, our approach improves SM by 17.6\%, IFN by 4.0\%, and NED by 42.8\%. However, TF-IDF + Soft outperforms our method in ICP on two benchmarks, achieving a 40.7\% average improvement.

\begin{table}[t]
  \centering
  \footnotesize
  \begin{tabular}{lrrrr}
  \toprule
  \textbf{Method} & \textbf{SM $\uparrow$} & \textbf{ICP $\downarrow$} & \textbf{IFN $\downarrow$} & \textbf{NED $\downarrow$} \\
  \midrule
  \multicolumn{5}{c}{\textbf{JPetStore}}\\
  \midrule
  UniXcoder + Hard & 0.097 & 0.407 & 1.733 & 0.185 \\
  UniXcoder + Soft (ours) & \textbf{0.289} & \textbf{0.120} & \textbf{1.470} & \textbf{0.025} \\
  \midrule
  \multicolumn{5}{c}{\textbf{DayTrader}}\\
  \midrule
  UniXcoder + Hard & 0.112 & 0.442 & 1.353 & 0.583 \\
  UniXcoder + Soft (ours) & \textbf{0.455} & \textbf{0.082} & \textbf{1.000} & \textbf{0.159} \\
  \midrule
  \multicolumn{5}{c}{\textbf{AcmeAir}}\\
  \midrule
  UniXcoder + Hard & 0.180 & 0.361 & 2.111 & \textbf{0.044} \\
  UniXcoder + Soft (ours) & \textbf{0.304} & \textbf{0.093} & \textbf{1.000} & 0.059 \\
  \midrule
  \multicolumn{5}{c}{\textbf{Plants}}\\
  \midrule
  UniXcoder + Hard & 0.277 & 0.453 & 3.042 & 0.095 \\
  UniXcoder + Soft (ours) & \textbf{0.629} & \textbf{0.080} & \textbf{0.938} & \textbf{0.000} \\
  \bottomrule
  \end{tabular}
  \caption{Comparison between UniXcoder + Hard and UniXcoder + Soft variants. \textbf{Bold} indicates better performance for each metric.}
  \label{tab:soft-hard}
\end{table}

\begin{table}[t]
  \centering
  \footnotesize
  \begin{tabular}{lrrrr}
  \toprule
  \textbf{Method} & \textbf{SM $\uparrow$} & \textbf{ICP $\downarrow$} & \textbf{IFN $\downarrow$} & \textbf{NED $\downarrow$} \\
  \midrule
  \multicolumn{5}{c}{\textbf{JPetStore}}\\
  \midrule
  TF-IDF + Soft & \textbf{0.531} & 0.121 & 1.532 & 0.132 \\
  UniXcoder + Soft (ours) & 0.289 & \textbf{0.120} & \textbf{1.470} & \textbf{0.025} \\
  \midrule
  \multicolumn{5}{c}{\textbf{DayTrader}}\\
  \midrule
  TF-IDF + Soft & 0.211 & \textbf{0.031} & 1.200 & \textbf{0.158} \\
  UniXcoder + Soft (ours) & \textbf{0.455} & 0.082 & \textbf{1.000} & 0.159 \\
  \midrule
  \multicolumn{5}{c}{\textbf{AcmeAir}}\\
  \midrule
  TF-IDF + Soft & 0.251 & \textbf{0.087} & 1.143 & \textbf{0.054} \\
  UniXcoder + Soft (ours) & \textbf{0.304} & 0.093 & \textbf{1.000} & 0.059 \\
  \midrule
  \multicolumn{5}{c}{\textbf{Plants}}\\
  \midrule
  TF-IDF + Soft & \textbf{0.792} & 0.087 & \textbf{0.800} & 0.027 \\
  UniXcoder + Soft (ours) & 0.629 & \textbf{0.080} & 0.938 & \textbf{0.000} \\
  \bottomrule
  \end{tabular}
  \caption{Comparison between TF-IDF + Soft and UniXcoder + Soft variants. \textbf{Bold} indicates better performance for each metric.}
  \label{tab:unixcoder-tfidf}
\end{table}

\begin{figure*}[t]
    \centering
    \includegraphics[width=0.8\textwidth]{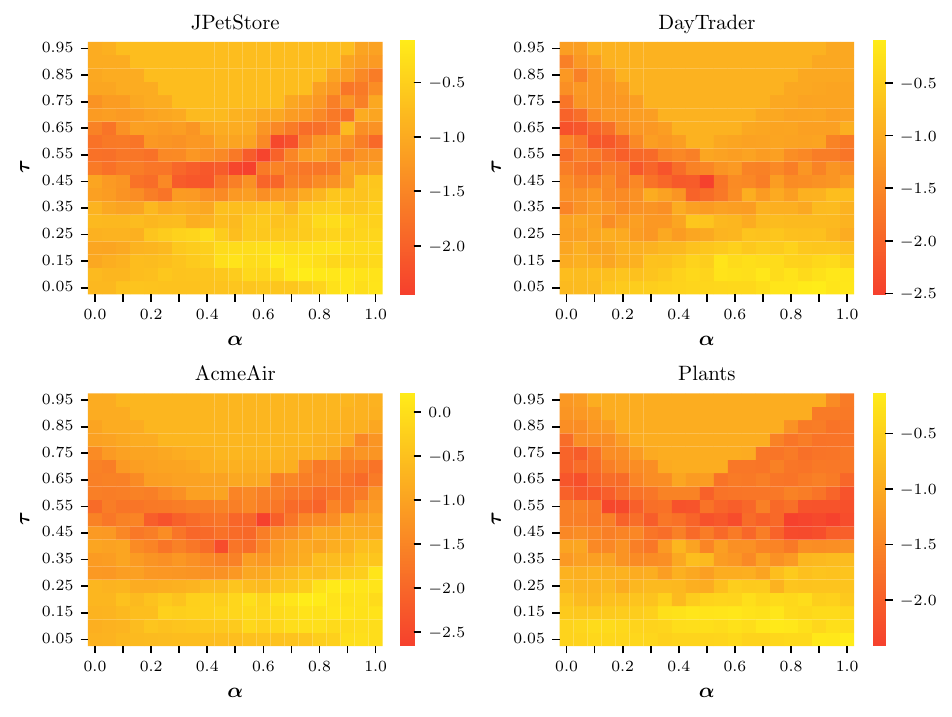}
    \caption{\textbf{Hyperparameter Sensitivity Analysis of Mo2oM Using QSCORE.} This figure illustrates the impact of $\alpha$ and $\tau$ on QSCORE across benchmarks, showing optimal configurations.}
    \label{fig:heatmap-plots}
\end{figure*}

\paragraph{Effects of Hyperparameters}
We investigate the sensitivity of our Mo2oM model to hyperparameters $\alpha$ (from Equation~\ref{eq:f-combined}) and $\tau$. The parameter $\alpha \in [0, 1]$ balances the influence of structural and semantic features, and the parameter $\tau \in (0, 1)$ governs the strictness of the partitioning strategy, determining how classes are grouped into microservices. To assess the impact of $\alpha$ and $\tau$ on the quality of the extracted microservices, we evaluate their effects across a range of values: $\alpha$ from 0 to 1 and $\tau$ from 0.05 to 0.95, both in increments of 0.05. For this analysis, we introduce a composite quality score, QSCORE, which aggregates normalized evaluation metrics as follows:
\begin{equation}
    \text{QSCORE} = \widetilde{\text{SM}} - \widetilde{\text{ICP}} - \widetilde{\text{IFN}} - \widetilde{\text{NED}},
\end{equation}
where $\widetilde{\text{M}}$ represents the normalized versions of metric $\text{M}$. Figure~\ref{fig:heatmap-plots} presents heatmaps of QSCORE values for different $\alpha$ and $\tau$ configurations across four benchmarks. In these heatmaps, brighter colors (yellow) indicate higher QSCORE values, reflecting superior performance. A consistent pattern emerges: a bright yellow region at the bottom-right of each heatmap suggests optimal results with higher $\alpha$ (semantic-heavy weighting) and lower $\tau$ (more permissive threshold). Specifically, $\alpha$ values between 0.6 and 1 yield the best outcomes, indicating that weighting semantic features more heavily ($\alpha > 0.5$) enhances microservice decomposition. This may reflect UniXcoder's ability to capture more meaningful class relationships than structural dependencies alone.
For $\tau$, optimal performance occurs within 0.1 to 0.3, while intermediate values (0.35 to 0.75) underperform. Lower $\tau$ values relax membership criteria, enabling classes to join closely related microservices, potentially forming larger, cohesive units. Higher $\tau$ values impose stricter criteria, producing smaller, highly modular services, but possibly compromising system coherence and producing disproportionate microservices.

\subsection{Discussion}

\paragraph{Flexibility of Mo2oM}
Effective migration to microservices involves selecting the right number of microservices: a larger number enhances deployment agility and fault isolation but also increases operational overhead. As a result, organizations with high engineering and infrastructure budgets benefit from a larger number of microservices, whereas smaller teams typically constrain microservices count to conserve resources. Mo2oM allows for specifying the target number of microservices up front, so it flexibly scales to both resource‑rich and resource‑constrained scenarios. In contrast, methods such as HyDec~\cite{sellami2022hierarchical} handle the service count internally, preventing users from adjusting the decomposition to match varying budgets. Building on our prior results, which demonstrated that Mo2oM consistently achieved superior median values across all baseline methods over a range of microservice counts, we further assess its adaptability under explicit budget constraints. We categorize budgets as "low," "medium," and "high" by uniformly partitioning the interval \([2, N/2]\) (stepping by two) to define the number of clusters hyperparameter values for each budget. Table~\ref{tab:flexibility} reports four benchmark applications evaluated under these three budgets. Across all benchmarks and budget levels, Mo2oM maintains stable results on all metrics. This invariance underscores Mo2oM’s ability to deliver consistently high‑quality decompositions, whether teams are constrained to very few microservices or capable of supporting many.

\begin{figure*}[t]
    \centering
    \includegraphics[width=0.95\linewidth]{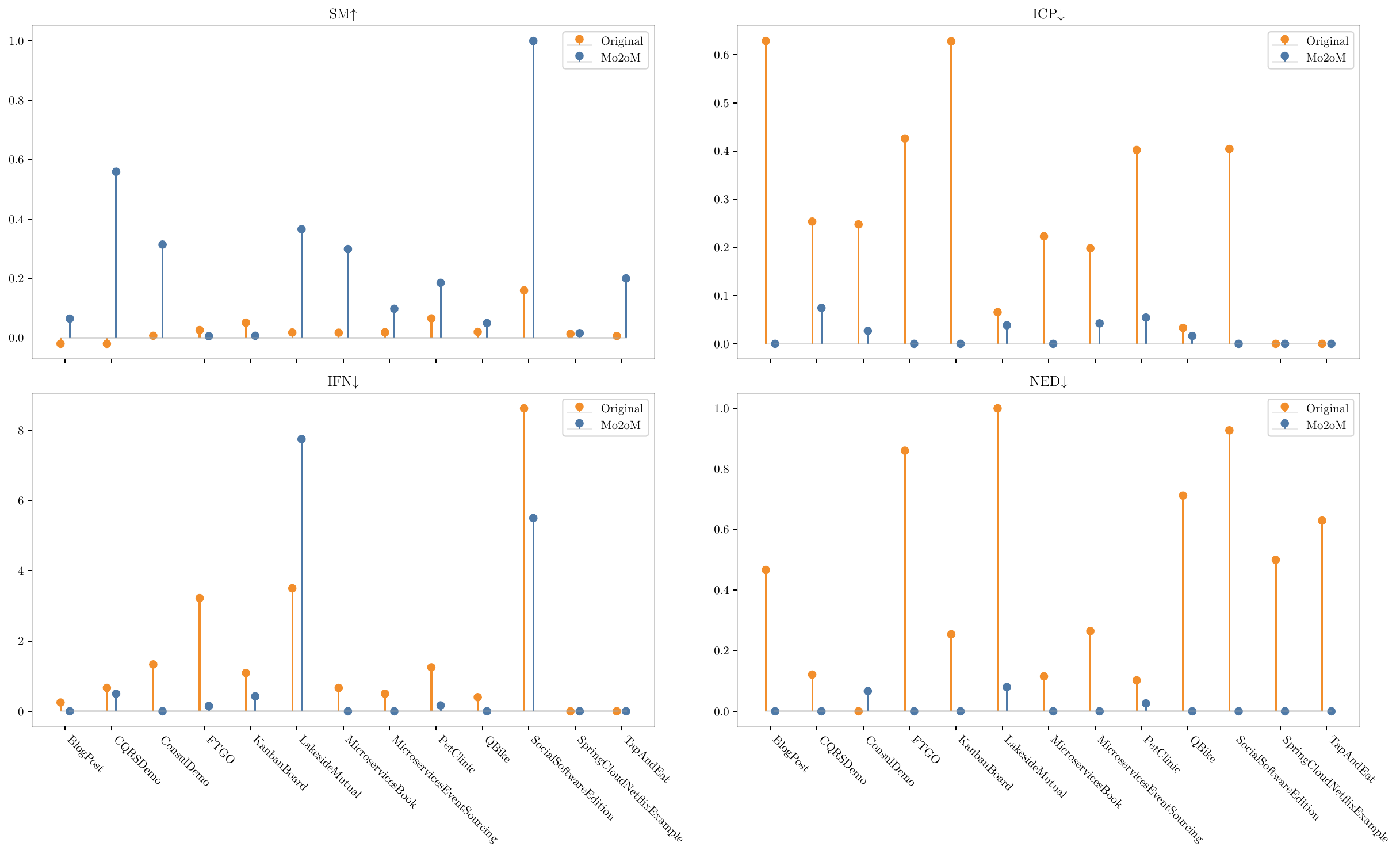}
    \caption{Quality metrics for original versus Mo2oM‐refactored decompositions across thirteen real‑world microservices applications.}
    \label{fig:fine-tuning}
\end{figure*}

\begin{table}[t]
\centering
\begin{tabular}{llrrrr}
\toprule
\textbf{Benchmark} & \textbf{Budget} & \textbf{SM $\uparrow$} & \textbf{ICP $\downarrow$} & \textbf{IFN $\downarrow$} & \textbf{NED $\downarrow$} \\
\midrule
\multirow{3}{*}{\textbf{JPetStore}} & Low & 0.328 & 0.095 & 3.175 & 0.000 \\
 & Medium & 0.295 & 0.080 & 1.185 & 0.161 \\
 & High & 0.237 & 0.184 & 0.986 & 0.059 \\
\midrule
\multirow{3}{*}{\textbf{DayTrader}} & Low & 1.595 & 0.055 & 2.559 & 0.109 \\
 & Medium & 0.497 & 0.072 & 1.065 & 0.160 \\
 & High & 0.426 & 0.149 & 0.601 & 0.211 \\
\midrule
\multirow{3}{*}{\textbf{AcmeAir}} & Low & 0.213 & 0.091 & 2.214 & 0.008 \\
 & Medium & 0.329 & 0.116 & 0.889 & 0.078 \\
 & High & 0.255 & 0.133 & 0.462 & 0.108 \\
\midrule
\multirow{3}{*}{\textbf{Plants}} & Low & 0.812 & 0.036 & 3.375 & 0.000 \\
 & Medium & 0.513 & 0.080 & 0.938 & 0.000 \\
 & High & 0.563 & 0.120 & 0.267 & 0.042 \\
\bottomrule
\end{tabular}
\caption{Performance of Mo2oM Under Different Budget Constraints. This table showcases adaptability under varying service counts, reporting key metrics.}
\label{tab:flexibility}
\end{table}

\paragraph{Refactoring Low‑Quality Microservices}
Microservice decomposition methods like Mo2oM are not only capable of migrating monolithic systems to microservices-based architectures but also excel at refactoring existing microservices deployments. To validate this, we applied Mo2oM to a suite of thirteen real‑world microservices applications drawn from established benchmarks~\cite{imranur2019curated, sellami2022hierarchical}. For each application, we compared the original decomposition against the Mo2oM‑generated decompositions.
As depicted in Figure~\ref{fig:fine-tuning}, the baseline decompositions of these applications suffer from near‑zero or negative structural modularity, excessive inter‑service coupling, and poorly balanced service sizes. By contrast, Mo2oM consistently produces decompositions with substantial improvements over all metric values. This demonstrates that even “ground‑truth” microservices designs can be meaningfully improved by Mo2oM’s automated approach. Consequently, Mo2oM provides a practical solution for systematically refactoring low‑quality microservices architectures.

\section{Conclusion}\label{conclusion}

While microservice extraction has become a cornerstone in modernizing monolithic systems, existing techniques largely rely on hard clustering, failing to account for the overlapping service boundaries observed in real-world architectures. We argue that supporting soft, overlapping class-to-service mappings better captures cross-cutting dependencies and enhances modularity. To this end, we propose Mo2oM, a novel framework that formulates microservice extraction as a soft clustering problem. Mo2oM integrates large language model-based semantic embeddings with structural dependency features and applies NOCD to generate probabilistic service memberships. Through extensive experiments across diverse monolithic applications, we demonstrate that Mo2oM consistently improves structural modularity, reduces inter-service communication, and balances microservice sizes, outperforming state-of-the-art baseline models. Ablation studies confirm the value of our LLM-based semantic modeling and soft clustering approach. Overall, Mo2oM offers a flexible, empirically grounded solution for producing high-quality microservice decompositions suitable for both migration and refactoring of existing microservices-based systems.

\paragraph{Limitations and Future Work}
 Mo2oM’s soft clustering approach introduces additional maintenance overhead due to repeated class definitions, complicating version control and deployment coordination. As future work, we will investigate automated refactoring and dynamic version‑tracking techniques to consolidate these overlaps across microservices and mitigate maintenance costs without compromising modularity.

\bibliography{aaai2026}

\end{document}